# A Novel Deep Learning based Model to Defend Network Intrusion Detection System against Adversarial Attacks

**Khushnaseeb Roshan**
Department of Computer Science
Aligarh Muslim University
Aligarh, India
kroshan@myamu.ac.in

**Aasim Zafar**
Department of Computer Science
Aligarh Muslim University
Aligarh, India
azafar.cs@amu.ac.in

**Shiekh Burhan Ul Haque**
Department of Computer Science
Aligarh Muslim University
Aligarh, India
sbuhaque@myamu.ac.in

*Abstract*— Network Intrusion Detection System (NIDS) is an essential tool in securing cyberspace from a variety of security risks and unknown cyberattacks. A number of solutions have been implemented for Machine Learning (ML), and Deep Learning (DL) based NIDS. However, all these solutions are vulnerable to adversarial attacks, in which the malicious actor tries to evade or fool the model by injecting adversarial perturbed examples into the system. The main aim of this research work is to study powerful adversarial attack algorithms and their defence method on DL-based NIDS. Fast Gradient Sign Method (FGSM), Jacobian Saliency Map Attack (JSMA), Projected Gradient Descent (PGD) and Carlini & Wagner (C&W) are four powerful adversarial attack methods implemented against the NIDS. As a defence method, Adversarial Training is used to increase the robustness of the NIDS model. The results are summarized in three phases, i.e., 1) before the adversarial attack, 2) after the adversarial attack, and 3) after the adversarial defence. The Canadian Institute for Cybersecurity Intrusion Detection System 2017 (CICIDS-2017) dataset is used for evaluation purposes with various performance measurements like f1-score, accuracy etc.

I. INTRODUCTION

Machine Learning (ML) and Deep Learning (DL) based algorithms are widely famous and extensively adopted in various sectors like Transportation [1], Healthcare [2], Image and Speech Recognition [3]–[5], Machine Translation [5], Network Intrusion Detection Systems (NIDS) [6], Cybersecurity [7]–[11] and much more. The tremendous growth of ML/DL based systems becomes possible due to the ease of affordable computational power, such as cloud services, and multiple GPU and TPU support, which leads towards promising results for future automation. The research community has been working to improve the efficiency of ML and DL algorithms in terms of various performance metrics for more than a decay [12] [13]. However, the generalization and robustness capability of the ML and DL algorithms cannot be ignored in today's era. This phenomenon includes the model's ability to deal with adversarial cyber attacks.

ML and DL methods are vulnerable to adversarial attacks, and these are intentionally crafted inputs (perturbations examples) to the models and mislead the system into producing incorrect results. Adversarial examples are the biggest vulnerability of the ML and DL algorithms, hence rendering its adoption in mission-critical applications such as NIDS, Streaming and Online ML/DL learning algorithms. Evasion, Extraction, Poisoning, and Inference are the four types of adversarial machine learning. In the case of NIDS, the Evasion attack led the model to misclassify the malicious network traffic as benign. The Poisoning attack aims to corrupt the NIDS model by inserting adversarial points during its training phase that cause the model to act in a way that is advantageous to the malicious user. The other type of attack is Extraction, where the malicious actor gathers information about the learning algorithms and their parameters to rebuild the same model and later uses it to attack and learn the targeted system. In the Inference attack, the malicious actor attempts to analyze the dataset information on which the learning system is trained or tested without accessing it. The black-box and white-box approaches are the two categories of adversarial machine learning. In the white-box approach, the malicious actor knows the learning algorithm; however, in the black-box method, only little or no information is known about the learning algorithm.

Adversarial machine learning is extensively explored in the unconstraint domain (e.g. image and object classification and recognition); however, it is less explored in constrained areas. In the case of the unconstrained domain, the adversary can fully exploit the features or pixels of the object/image. But in a constrained domain, the situation is different as 1) features may be correlated with each other, 2) it can be binary, continuous or categorical, and 3) some feature values can not be changed by the adversary and remain fixed. All these factors make it unclear "whether the constrained domain is less vulnerable to adversarial examples?". This hypothesis is tested by Sheatsley et al. [14]. The authors concluded that generated misclassification rate is greater than 95% when experimented with two algorithms, adaptive JSMA and histogram sketch generation.

In this research work, we empirically analyzed the effects of four powerful adversarial attack algorithms: FGSM [15], JSMA [16], PGD [17] and C&W [18] over the DL-based NIDS. We further studied one of the most powerful defence methods to safeguard NIDS from adversaries by applying the adversarial training defence mechanism.

The overall research is organized as follows: Section II discusses the related study, and Section III explains the proposed approach. Section IV discusses the experimental results, and Section V concludes the overall research work.

## II. RELATED WORK

Adversarial machine learning is a common area of machine learning and computer security. Szegedy et al. [19] first discovered the intriguing properties of Neural Networks (NN) in 2014, which revealed that the NN are vulnerable to adversarially crafted inputs. The authors empirically proved this by generating adversarial inputs with the box-constrained optimization-based algorithm. Three datasets, namely, MNIST, ImageNet and approx 10M image samples from Youtube, are utilized with different NN architectures and hyperparameters settings. Since then, the research community has been exploring and searching for new methods for adversarial machine learning.

Wang [20] explored adversarial machine learning algorithms in a supervised NIDS. The author explored the effect of four methods, namely FGSM, JSMA, DeepFool, and C&W, to analyse the impact of the top features that are altered and contribute to generating adversarial examples. Through extensive experimentation, the reduced confidence score is evaluated with various performance measures like f1-score, accuracy, precision, AUC value etc. However, the research work can be further extended to include recent defence strategies to secure NIDS from adversarial attacks. Pawlicki et al. [21] studied the impact of adversarial machine learning on ML and DL based NIDS. The authors used DL, Evolutionary Computation and Monte Carlo methods to generate perturbed examples to fool NIDS into producing incorrect predictions. The latest CICIDS-2017 dataset is used for evaluation purposes over the five ML and DL algorithms, namely, ANN, Random Forest, AdaBoost, SVM and KNN.

Guo et al. [22] studied adversarial machine learning in cybersecurity and compared the generation of adversarial examples in computer vision and NIDS. The authors applied the Basic Iterative Method (BIM) [23], it is the extension of the FGSM [15] method with a multiple-step size applied iteratively. The authors built two models, the first was the target, and the second was the substitute model. The KDDCUP-99 and CICIDS-2018 two benchmark datasets are used for experimentation. The BIM method generates adversarially perturbed examples to attack the target system. However, the research work can be extended to explore more complex adversarial attack strategies as well as their defences to increase the robustness of the model. Qureshi et al. [24] proposed a novel algorithm, Random Neural Network based Adversarial NIDS (RNN-ADV), along with the JSMA algorithm. JSMA algorithm is more efficient in terms of resource utilization as it changes only few features to create perturbed examples. The NSL-KDD dataset is used for experimentation, and a comparison is made between the MLP and the proposed RNN-ADV algorithm to show its effectiveness.

Alhajjar et al. [25] studied adversarial machine learning in NIDS and explored Genetic, Particle Swarm Optimization Evolutionary algorithms and Generative Adversarial Networks (GAN) along with Monte Carlo simulation for adversarial examples generation. Two publicly available NSL-KDD and UNSW-NB15 datasets are used for extensive experimentation over the eleven ML algorithms with evasion rate as performance evaluation measurement. The authors also discussed the transferability phenomena of the ML and DL models, which implies that an input meant to fool one model can cause a similar behaviour to occur in a different model.

Usama et al. [26] explored the use of GAN for both adversarial attacks and defence mechanisms in ML and DL based NIDS. The authors evaluated eight ML/DL NIDS classifiers: DNN, Random Forest, Logistic Regression, Support Vector Machine, K-Nearest Neighbour, Decision Trees, Gradient Boosting, and Naıve Bayes, techniques as black-box IDS. And as a defence strategy, adversarial training with GAN is used to mitigate the effect of adversarial attacks and to increase the robustness of the NIDS model. Furthermore, to preserve the network traffic functional behaviour, the complete features set is divided into functional and non-functional attributes. Furthermore, based on the non-functional sets, adversarial examples are created. The KDD CUP-99 dataset is used, and the results are evaluated based on accuracy, precision, recall and f1-score.

The extensive literature review shows that not much work has been explored in adversarial machine learning under the constrained domain. Hence, in this research article, we have explored and examined the various adversarial attack and defence methods in the network intrusion detection domain.

## III. PROPOSED APPROACH

This section describes the proposed approach starting from dataset selection, pre-processing, and model building, followed by adversarial attacks and defence methods.

*A. Dataset and Pre-processing*

The subset of the CICIDS-2017 [27] dataset is used for experimentation. It is developed by the Canadian Institute for Cybersecurity and encapsulated into five files from Monday to Friday. The Monday file contains only benign data, and the remaining files include both benign and malicious network traffic. The dataset is publicly available in PCAP, Generated Labelled Flows and CSV file formats for ML and DL applications. It consists of seventy-nine features, which are divided into statistical measurements such as average packet flow, standard deviation, min-max packet counts, and other packet flow and packet size distributions etc. The CICIDS-2017 dataset is up to date and contains a variety of the latest insider and outsider attacks with fourteen classes, such as DDoS, DoS, Heartbleed, Bot, Infiltration etc. Hence this dataset would be the right choice to efficiently evaluate the NIDS model in both normal and adversarial attack situations. The dataset is pre-processed to remove any null and infinity values and divided into training, validation and testing sets with the scikit-learn ML function.

*B. Deep Learning based NIDS Model*

Deep Neural Networks (DNN) is an Artificial Neural Network (ANN) which consists of one input layer, one output layer and one or more hidden layers. Each layer consists of artificial neurons and activation functions. The number of hidden layers and neurons may vary depending on the complexity of the problem to be solved. In this research work, we used a supervised DNN algorithm to build the NIDS system with optimal hyperparameters. The selection of the optimal structure of DNN and the other parameters, such as learning rate, activation function etc., are based on random search [28] algorithms which converge faster compared to grid search algorithms, with semi-optimal parameter sets [29]. The NIDS model is trained and validated on the subset of the CICIDS-2017 dataset with 50 epochs, as shown in Fig. 1. Fig. 2 represents the conceptual architecture of the proposed approach. The initial phase consists of dataset selection, pre-processing and splitting into training and testing sets. The next phase is model training and testing. Adversarial attacks are encountered during the testing phase and detected by the administrator. As a defence mechanism the Adversarial Training is implemented to safeguard NIDS against adversarial attacks.

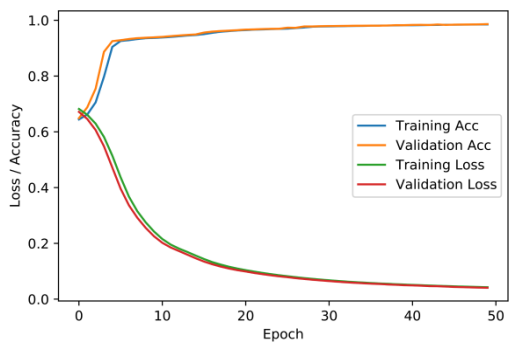

Fig. 1. NIDS model training and validation accuracy on CICIDS-2017

*C. Adversarial Examples Generations*

In this research study, FGSM, JSMA, PGD and C&W, four powerful adversarial attack generation algorithms, are utilized to generate perturbed examples to fool DL based NIDS model.

Goodfellow et al. [15] proposed the FGSM method to generate adversarial examples based on the gradient sign method using backpropagation. The algorithm is based on the optimization of Lp norm (distance). It is an untargeted attack approach used to obtain max-norm constrained perturbation η expressed in Equation (1). Here θ represents the model parameter, *x* is the input vector to the model and *y* is the associated label of the input and $J(\theta,x,y)$ is the cost function. FGSM rapidly generate perturbation samples with a small noise parameter $\epsilon$.

$$\eta = \epsilon \; sign\left(\nabla x \, J(\theta, x, y)\right) \quad (1)$$

Papernot et al. [16] proposed JSMA method based on the Jacobian Matrix, which aims to calculate the forward derivative of the cost function *f(x)*. JSMA algorithm is more efficient as it iteratively calculates the saliency map hence leading to identifying the most significant input features that contribute more to model predictions and triggers large variations. The problem can be formulated as shown in Equation (2). Here *x* is the input feature vector to the NN-based model.

$$Jf(x) = \frac{\partial f(x)}{\partial x} = \left[\frac{\partial fj(x)}{\partial xi}\right] i \times j \quad (2)$$

PGD [17] adversarial examples generation algorithms are based on the first-order $L_\infty$ norm that searches iteratively for the perturbation and optimizes the saddle point (min-max) formulation.

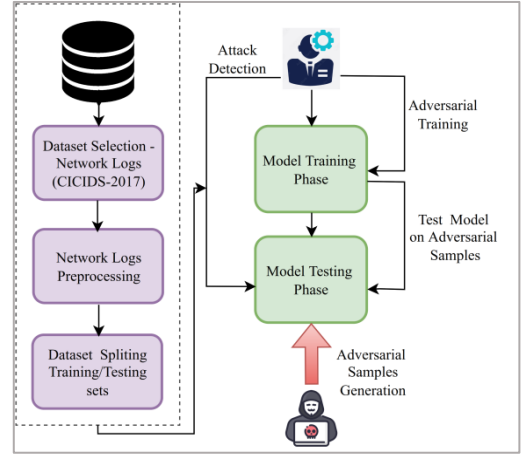

Fig. 2. Conceptual architecture of adversarial attack and defence methods

Madry et al. [17] addressed two main issues of adversarial machine learning. The first is "generating strong adversarial examples with only small noise," and the second is "model training should be done in such a way that no perturbed examples are possible or difficult to find by adversaries". The problem is formulated as in Equation (3).

$x0adv = x,$

$$x(n+1) = Clipx, \epsilon \{xn + \epsilon \; sign(\nabla xJ(\theta,x,y))\} \quad (3)$$

Carlini & Wagner (C&W) is one of the most powerful adversarial example generation methods [18] that makes perturbed examples imperceptible. It is based on Lp norms (0, 2 and inf) and can defeat the defensive distillation method [30]. The core of the objective function is shown in Equation (4). Here, δ is the minimum noise, *f( )* is the objective function, and c is the constant obtained from the modified binary search.

$$\min \delta \; ||\delta||p + c.f(x+\delta)$$

$$f(x') = \max(\max\{Z(x')i : I \neq t\} - Z(x')t, -K \quad (4)$$

*D. Adversarial Defence Strategy*

The adversarial defence training method [31] is used to robust the model in handling adversarial attacks. It is one of the most effective and intuitive defence methods in which the augmented inputs are inserted into the model during the training phase. The model is re-trained with adversarially generated samples with all four approaches (FGSM, JSMA, PGD, and C&W) to increase diversity during the training, thus improving the defensive capability against the transferred adversaries, single and multi-steps attacks as well. The complete methodology is summarized in Algorithm 1.

**Algorithm 1** : Model evaluation under Attack and Defence methods

> **Input**: test_data, original_trained_model M
> **Output**: classification_report C_R, ROC
> # list of variables
> true_label ← test_data(true_label)
> adversarial_algo←{fgsm, jsma, pgd, c&w}
> adv_list ←{}
>
> # generate adversarial examples
> **for** algo **in** adversarial_algo **do**:
>  adv = generate_adversarial_samples(test_data, M, algo)
>   adv_list.append(adv)
> **end for**
>
> # evaluate the model under adversarial attack
> **for** lst **in** adv_list **do**:
>   predict ←M.predict(lst)
>   C_R ← classification_report (predict, true_label)
>   ROC ← roc_curve (predict, true_label)
>   **return** CR CM ROC
>  **end for**
>
> # Build and evaluate defence model
> defence_model ← AdversarialTrainer (M, adv_list, algo_param)
> **for** lst **in** adv_list **do**:
>   predict ←defence_model.predict(lst)
>   C_R ← classification_report (predict, true_label)
>   ROC ← roc_curve (predict, true_label)
>   **return** C_R ROC
>  **end for**

*E. Evaluation Metrics*

The performance evaluation of the NIDS model is demonstrated with accuracy, precision, recall and f-score in all three scenarios, i.e., under normal conditions, under adversarial attacks and after the adversarial defence. Equations (5) – (8) are the performance measurement metrics formulated over the binary classification terms: True Positive, True Negative, False Positive and False Negative cases.

$$\text{Accuracy (ACC)} = \frac{TP + TN}{TP + TN + FP + FN} \quad (5)$$

$$\text{Precision (P)} = \frac{TP}{TP + FP} \quad (6)$$

$$\text{Recall (R)} = \frac{TP}{TP + FN} \quad (7)$$

$$F1 - \text{Score(F)} = \frac{2 \times R \times P}{R + P} \quad (8)$$

## IV. EXPERIMENT AND EVALUATION

This experimentation is conducted on Windows-11 OS, Core-i7 CPU, 16-GB RAM, 500-GB SSD, Python 3.7 and other supportive libraries such as pandas, scikit-learn, IBM ART-toolbox.

The initial DL-based NIDS model is the supervised architecture with three hidden layers consisting of 50, 30 and 10 neurons in each layer. The learning rate of 0.001 is used, within the hidden layers relu activation function is used, and the sigmoid activation function is in the output layer. The NIDS model is trained with 50 epochs on the training data set with a validation-split parameter set to 0.25 for model validation. Finally, the model is trained with 50 epochs. The obtained accuracy, precision, recall and f1-Score of the model under normal situations (without adversarial attack) are 98.54%, 98.56%, 98.54%, and 98.54%, as shown in Table 1. The obtained AUC score of the NIDS model is 98.551, as shown in Fig. 3.

TABLE I. RESULTS OF THE NIDS MODEL AFTER ADVERSARIAL ATTACK AND DEFENCE

|  | In (%) | FGSM (%) | | JSMA (%) | |
| --- | --- | --- | --- | --- | --- |
|  | Before Attack | After Attack | After Defence | After Attack | After Defence |
| ACC | 98.54 | 57.95 | **98.7** | 66.58 | **98.47** |
| P | 98.56 | 66.93 | **98.72** | 77.66 | **98.51** |
| R | 98.54 | 57.95 | **98.7** | 66.58 | **98.47** |
| F | 98.54 | 52.54 | **98.7** | 66.33 | **98.47** |
|  | In (%) | PGD (%) | | C&W (%) | |
|  | Before Attack | After Attack | After Defence | After Attack | After Defence |
| ACC | 98.54 | 56.81 | **98.68** | 61.74 | **71.56** |
| P | 98.56 | 66 | **98.71** | 74.82 | **80.35** |
| R | 98.54 | 56.81 | **98.68** | 61.74 | **71.56** |
| F | 98.54 | 51.47 | **98.68** | 57.11 | **69.86** |
| ACC-Accuracy, P-Precision, R-Recall, F-F1-score | | | | | |

The NIDS model is evaluated under the adversarial attack cases with FGSM, JSMA, PGD and C&W four evasion adversarial algorithms. The aim is to reduce the performance of the model. The obtained reduced accuracy, precision, recall and f1-score under the FGSM, JSMA, PGD and C&W adversarial attack generation algorithms are (57.95%, 66.93%, 57.95%, 52.54%), (66.58%, 77.66%, 66.58%, 66.33% ), (56.81%, 66%, 56.81%, 51.47%) and (61.74%, 74.82%, 61.74%, 57.11), respectively. The results are also evaluated in terms of the ROC curve, as shown in Fig. 3. The resulting reduced AUC score under the four algorithms are 59.232, 68.037, 58.485, and 63.41, respectively. It is clearly visible that the performance of the NIDS model is almost reduced to half in all four cases.

As discussed previously, adversarial training is one of the most effective and intuitive approaches for the adversarial vulnerability of the ML and DL models. Hence to increase the robustness of the NIDS model and its defensive capability, the re-training is done along with adversarial perturbation examples generated with FGSM, JSMA, PGD and C&W methods. The built defensive model is again evaluated over the perturbated examples. As a result, the model performance is significantly improved. The resulting improved accuracy, precision, recall and f1-Score, under the FGSM, JSMA, PGD and C&W scenarios are illustrated in Table 1.

A Novel Deep Learning based Model to Defend Network Intrusion Detection System against Adversarial Attacks

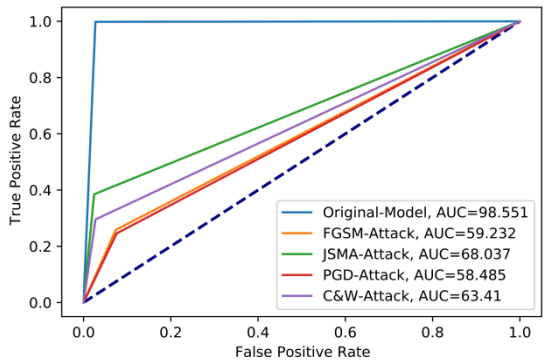

Fig. 3. NIDS ROC curve before and after the adversarial attack

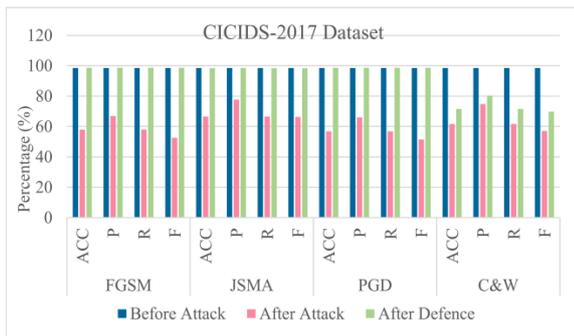

Fig. 4. NIDS performance comparison before attack, after attack and after defence

Fig. 4 graphically demonstrates the results for clear visualization under all three conditions, namely, before the attack, after the adversarial attack and after the adversarial defence. C&W is regarded as one of the most potent adversarial sample generation algorithms, but after experimentation, we see quite similar effects in terms of reducing confidence score compared to other algorithms (FGSM, JSMA and PGD). After the adversarial defence implementation, significant improvement is achieved under FGSM, JSMA and PGD. However, in C&W method, the maximum reach is up to 80.35% only in terms of precision. Hence, future work may involve investigating and increasing model robustness in C&W adversarial attacks. The adversarial transferability is another interesting phenomenon of ML and DL models in which the adversarial perturbation examples also similarly impact other models having different architecture and parameter settings.

## V. CONCLUSION

This research study concluded that no domain weather constraints or un-constraints are secure from adversarial attacks. The same idea is demonstrated with the FGSM, JSMA, PGD and C&W, four powerful adversarial algorithms to fool the DL-based NIDS model to misclassify the benign samples into anomalies and vice-versa. The NIDS model is first evaluated under a normal situation. The achieved accuracy and f1-score of the model is 98.54%. Later, the same model is evaluated under the adversarial attack situation with FGSM, JSMA, PGD, and C&W methods. As a result, the accuracy, f1-score, and AUC value have significantly reduced. The adversarial defence approach is used to mitigate the effect of adversarial attacks and to improve the robustness and confidence score of the model.

After adversarial training, the improved accuracy under FGSM, JSMA, PGD and C&W are 98.7%, 98.47%, 98.68%, and 71.56%, respectively. In future work, the proposed approach could be extended with ML and DL architecture and recent intrusion detection datasets to see the impact of adversarial attack and defence methods.